\documentclass[fleqn,twocolumn, a4paper]{aa}

\usepackage{graphicx}
\usepackage{amsmath}
\usepackage{amssymb}
\usepackage{array}
\usepackage{subeqnarray}
\usepackage{natbib}
\bibpunct{(}{)}{;}{a}{}{, }
\usepackage[british]{babel}

\newcommand{\be}{\begin{equation}}
\newcommand{\ee}{\end{equation}}
\newcommand{\bea}{\begin{eqnarray}}
\newcommand{\eea}{\end{eqnarray}}
\newcommand{\nn}{\nonumber}
\newcommand{\beqs}{\begin{subeqnarray}}
\newcommand{\eeqs}{\end{subeqnarray}}
\newcommand{\pd}{\partial}

\def\pa{\shortparallel}

\def\grad{\vec{\nabla}}
\def\div{\vec{\nabla}\cdot}
\def\rot{\vec{\nabla}\times}

\begin{document}

\title{Magnetic fields from reionisation}

\author{Mathieu Langer\inst{1}
 \and Nabila Aghanim \inst{2}
\and Jean-Loup Puget\inst{2}}

   \offprints{M. Langer,\\ \email{mlanger@astro.ox.ac.uk}}

   \institute{Astrophysics - Denys Wilkinson Building - Keble Road - Oxford OX1 3RH - United Kingdom \\
\and
   Institut d'Astrophysique Spatiale - Universit\'e Paris-Sud - 91405 Orsay cedex - France}

   \date{Received: / Accepted: }

   \abstract{We present a complementary study to a new model for generating magnetic fields of cosmological interest. The driving mechanism is the photoionisation process by photons provided  by the first luminous sources. Investigating the transient regime at the onset of inhomogeneous reionisation, we show that magnetic field amplitudes as high as  $2\times 10^{-16}$ Gauss can be obtained within a source lifetime. Photons with energies above the ionisation threshold accelerate electrons, inducing magnetic fields outside the Str\"omgren spheres which surround the ionising sources. Thanks to their mean free path, photons with higher energies propagate further and lead to magnetic field generation deeper in the neutral medium. We find that soft $X$-ray photons could contribute to a significant premagnetisation of the intergalactic medium at a redshift of $z=15$.


   }
\authorrunning{M. Langer \& N. Aghanim \& J.-L. Puget}
   \titlerunning{Magnetic fields from reionisation}

   \maketitle


\section{Introduction}
Various observational techniques have been employed to detect the presence of magnetic fields in very different structures and on very different scales. On the largest scales, the measured amplitudes of the magnetic field are generally in the range of a few tenths microgauss to tens of microgauss, both in galaxies \citep[e.g.][]{becketal} and  on cluster scales \citep[][]{kron,CaTa}. Faraday rotation measurements have permitted to obtain an upper limit for the intergalactic magnetic field of $\left| B_\text{igm} \right| \lesssim 10^{-9}$ G, assuming a maximal field reversal scale of 1 Mpc \citep[][]{kron94}.

The origin of those magnetic fields remains largely unknown. It has been soon realised that cosmic  magnetic fields could not have been born directly with such high amplitudes. Had it been so, the structure formation history would have been drastically modified, since a field of $10^{-9}$ G on megaparsec scales is strong enough to induce nonlinear matter density fluctuations \citep[][ see also \citealt{kor}]{wasser}. Clearly then, cosmic magnetic fields must first  arise in the form of weak seeds, the amplitude of which is then amplified by some powerful mechanism. 

A plethora of models for the creation of cosmic magnetic fields is available in the literature, but none of the mechanisms envisaged seems completely satisfactory.  On the one hand, there is a large variety of models relying on the physics of the early universe, but the predicted seed amplitude is strongly model dependent, and spans consequently a rather large interval \citep[$10^{-65}$ G to $10^{-9}$ Gauss, typically - for recent reviews, see][]{graro, widrow, massimo}.

On the other hand, after recombination, most of the astrophysical
mechanisms proposed \citep[e.g.][]{joe,dw} are based on the Biermann
battery mechanism \citep[][]{battery}, first investigated to account
for stellar magnetism. Applications of the battery mechanism to
various cosmological contexts lead to seed fields of the order of
$10^{-20}-10^{-17}$ G on scales of a few kiloparsecs to a few hundreds
of kiloparsecs \citep[for the context of reionisation in particular, see][]{subra, gfz}. Since those models rely on the same
mechanism, namely differential acceleration of electrons (vs. ions)
due to thermal pressure, acting on similar scales, it is not
surprising that the range obtained for the initial magnetic field is
rather narrow (as compared to early universe models). Those fields are
usually thought to be sufficient to account for present day galactic
fields provided a subsequent amplification by galactic dynamo.

In \citet[][]{first}, we proposed an efficient magnetogenesis model
in which the driving force is the radiation emanating from
the first luminous objects.  Under the assumption of stationary
regime, we found that anisotropic and inhomogeneous radiation flux
leads, at the end of cosmological reionisation, to magnetic fields as 
high as $10^{-12} - 10^{-11}$ Gauss on protogalactic
scales. Moreover, we also found that magnetic fields on small scales
are strongly damped. The advantage of our model resides mainly in two
facts. First, the driving force is the radiation drag which
provides an acceleration of charged particles inversely proportional
to their mass cubed. As a consequence, electric currents develop
millions of times more efficiently as compared with Biermann battery
models which rely on thermal pressure. Second, the properties of the
magnetic field power spectrum are directly related to the shape of the
power spectrum of the matter density fluctuations. This guarantees
that large scale fields are favoured relative to small scale fields.

 In the present article, we extend our first model to the case of
the transient regime. More specifically, we focus on the generation of magnetic
fields at the onset of reionisation. The driving process for charge separation and acceleration is photoionisation itself. The formalism and the results are
given in Sects. \ref{model} and \ref{om}.  Discussion and conclusions are presented in
Sects. \ref{fin} and \ref{fin2}.

\section{Formalism}\label{model}
We study the generation of magnetic fields due to electric currents induced by the acceleration provided by radiation drag. The relevant equations are the usual Maxwell equations 
\beqs\label{max}
\div{\vec{E}} = 0  &  & \pd_t\vec{E} = c\rot{\vec{B}} - 4\pi \vec{j}\\
\div{\vec{B}} = 0  &  & \pd_t\vec{B} =-c\rot{\vec{E}}
\eeqs
where the electric current is $\vec{j} = n_eq_e(\vec{v}_e - \vec{v}_p)$,  
  $n_e$ being the electron density, $q_e$ the electron charge, $\vec{v}_{e(p)}$ the electron (proton) velocity, 
 and we have supposed  that the medium is globally neutral.
 We also consider the generalised Ohm's law,
\bea
\pd_t\vec{j} + \left(\vec{u}\cdot\grad\right)\vec{j} &=& \frac{\omega_p^2}{4\pi} \left[\vec{E} + \frac{\vec{u}\times\vec{B}}{c}\right] + \frac{q_e}{m_ec}\vec{j}\times\vec{B} \nn \\
& & -\nu_c\vec{j} + \nu_{\gamma}\vec{I},\label{ohms}
\eea
where $\omega_p$ is the electron plasma frequency,  $\nu_c$ is the electron-ion collision frequency, $m_e$ the electron mass, and $\vec{u}$ is the peculiar velocity field of the center of mass of the plasma. We included the radiation drag effects as a source term, 
\be
\nu_{\gamma} \vec{I}  \propto \frac{h\nu}{m_ec^2}\sigma_{\gamma}\vec{\phi},
\ee
defining the photon plasma interaction frequency by $\nu_{\gamma}=\sigma_{\gamma}\phi$  , where $h\nu$ is the photon energy, $\phi$ is the photon flux and $\sigma_{\gamma}$  the photon-electron interaction cross-section. The source current $I$ obviously depends on the ionisation state of the medium  (see Sect. \ref{om}). Initially, all but  the source term are zero, and we may explore the generation of the magnetic fields by considering linear terms only. We defer to Sect. \ref{fin} the discussion of nonlinear terms.
Equation (\ref{ohms}) 
shows three characteristic frequencies, $\omega_p\sim 5.64\times 10^{4}n_e^{1/2}\text{rad}\,\text{s}^{-1}$, $\nu_c \sim 50 n_e T^{-3/2}\text{s}^{-1}$ and $\nu_{\gamma}$. For illustration, at redshift $15$, when the universe is almost neutral, we have a  residual ionisation of $n_e \simeq 2.1\times 10^{-7} \,\text{cm}^{-3}$  \citep[see][ after \citeauthor{sss} \citeyear{sss}]{psh}, and the plasma frequency is $\omega_p \simeq 26 \,\text{rad}\,\text{s}^{-1}$, which is much larger than the electron ion collision rate, $\nu_c\sim 1.35 \times 10^{-5} \,\text{s}^{-1}$ (at a mean gas temperature $T\sim 0.85$ K, as redshifted  from $T\sim 4000$ K at $z=1100$ to $z=15$). Finally, estimating the photon flux at the hydrogen ionisation threshold as $\phi \sim 4.56 \times 10^{31}\,\text{cm}^{-2} \text{s}^{-1}$ (at a distance of $10^2$ kpc of  a  $10^{12} L_\odot$ source, see eq. (\ref{qf}) below), we obtain $\nu_\gamma \sim 3\times 10^7\, \text{s}^{-1}$, considering photon-electron Thomson scattering.

Thus, the development of electric currents from $\vec{j} = 0$ can be decomposed in three distinct time sequences. First, for times shorter than $\omega_p^{-1}$, 
only the source term is relevant. Considering it is constant in time,
we obtain
\beqs
\vec{j} &=&  \nu_{\gamma} \vec{I}  t\\
\vec{E} &=& -2\pi \nu_{\gamma} \vec{I} t^2 \\
\vec{B} &=& \frac{2\pi}{3} c \nu_{\gamma} \rot \vec{I} t^3.
\eeqs
So initially, the magnetic field grows quite quickly, but this regime is valid on a very short time only,
 i.e. for $t<\omega_p^{-1}$.

However, very soon $t$ reaches  $\omega_p^{-1}$, and in that case we need to consider the electric part of Lorentz force, in which case the system to solve is
\beqs\label{pla}
\pd_t\vec{j} &=& \nu_{\gamma} \vec{I} + \frac{\omega_p^2}{4\pi} \vec{E} \\
\pd_t\vec{E} &=& -4\pi \vec{j} +  c\rot{\vec{B}}\\
\pd_t\vec{B} &=& -c\rot\vec{E}.
\eeqs
Using Fourier and Laplace transforms, we obtain
\beqs\label{voici}
\vec{\tilde{E}}_k &=& -4\pi\nu_{\gamma}\frac{1-\cos{\left[\omega_{\tau} t\right]}}{\omega_{\tau}^2}\vec{\tilde{J}}_k \label{voici_te}\\
\vec{\tilde{j}}_k &=& \left[(1- \frac{\omega_p^2}{\omega_{\tau}^2})t+\frac{\omega_p^2}{\omega_{\tau}^2}\frac{\sin{\left(\omega_{\tau} t\right)}}{{\omega_{\tau}}}\right] \nu_{\gamma}\vec{\tilde{J}}_k \\
\vec{\tilde{B}}_k &=& - i\frac{4\pi\nu_{\gamma}c}{\omega_{\tau}^2}\left[ t -\frac{\sin{\left(\omega_{\tau} t\right)}}{\omega_{\tau}} \right]\vec{k}\times\vec{\tilde{J}}_k\label{mag-t}
\eeqs
where $\omega_{\tau} =\left(\omega_p^2 + c^2k^2\right)^{1/2} \sim \omega_p$ for scales $k\ll k_p =\omega_p/c\simeq 1.82\times 10^{-4}\,n_e^{1/2}\,\rm{m}^{-1}$ . Taking for instance the  residual ionisation after recombination at a redshift of $z\sim 15$, we have  $n_e \sim 2.1\times 10^{-7}\,\rm{cm}^{-3}$, and  the corresponding plasma scale is $k_p^{-1}\sim 1.2\times 10^{7}$ m.  After reionisation is complete, say at $z\sim 6$, we have $n_e \sim 8.1\times 10^{-5}\,\rm{cm}^{-3}$, and the plasma scale is even smaller. Therefore, for all scales of  interest, the characteristic pulsation is  $\omega_{\tau} =  \omega_p$.

Because of the intrinsic velocity distribution of the electrons, the 
oscillations shown in
eqs. (\ref{voici}) will vanish by Landau damping, essentially within a
few $\tau_p \sim \omega_p^{-1}$. This means that after a few plasma
time scales, the electric field  can be considered constant, the
electric current is vanishing (except perhaps on very small scales,
i.e. for $k\gg k_p$, where $\vec{j}$ continues to grow linearly), and
the magnetic field is increasing linearly with time.

Finally, the next time scale involved is the magnetic field diffusion
time, $\tau_D = \omega_p^2k^{-2}/(c^2\nu_c)$, until which the
evolution of $\vec{B}$ will be given (from eqs. \ref{mag-t}) by
\be
\vec{B} = 4\pi \frac{c\nu_\gamma}{\omega_p^2}\, t \,\rot\vec{I}\label{field}
\ee
on scales bigger than $k^{-1}_p$.

\section{Magnetic field seeds}\label{om}
We can now estimate the order of magnitude reached by the magnetic field in a source lifetime $t_\text{S}$. However, in the vicinity of an ionising source, the actual physical process leading to electron acceleration will depend on the ionisation state of the medium.

\subsection{The Str\"omgren sphere}
The extent of the ionised region around a luminous source depends on the recombination rate. If recombinations are negligible, the radius of the ionised sphere depends on the age $t_\text{S}$ of the source, 
\be
r_\text{S} = \left(\frac{3\dot{N}_\text{ph}t_\text{S}}{4\pi\langle n_\text{H}\rangle}\right)^{1/3},
\ee 
where $\dot{N}_\text{ph}$ is the source ionising photons emission rate and $\langle n_\text{H}\rangle$  is the mean hydrogen density.
In the opposite case, it is determined by the balance between recombination and emission of ionising photons, and its value is given by
\be\label{sr}
r_\text{S}  = \left(\frac{3\dot{N}_\text{ph}}{4\pi \alpha_\text{rec} C \langle n_\text{H}\rangle^2 } \right)^{1/3}, 
\ee
where
$\alpha_\text{rec} = 2.6\times 10^{-13}\ \text{cm}^3\text{s}^{-1}$ is
the radiative recombination coefficient at a gas temperature of
  $10^4$ K \citep[][]{humsea} and $C\equiv \langle
n_\text{H}^2\rangle/\langle n_\text{H}\rangle^2 >1$ is the clumping
factor.

The recombination time scale can be defined as \citep[e.g.][]{mhr}
\bea
t_\text{rec} &=& \left[1.17 \langle n_\text{H}\rangle \alpha_\text{rec} C\right]^{-1} \nn\\
&=& 1.26\times 10^8 C^{-1}\left(\frac{\Omega_b h^2}{0.024}\right)^{-1}\left(\frac{1+z}{16}\right)^{-3}  \text{yrs} \label{t-rec}
\eea
 Depending on its definition and on the model of structure
formation, very different values for $C$ are
obtained. \citet[][]{bnsl} for instance obtain values slightly above 3
at $z=15$, whereas calculations by  \citet[][]{patjoe} yield $C\sim
40$ at the same redshift. Also, results from simulations are quite
sensitive to numerical resolution, a point early emphasized by \citet[][]{go}. For instance, \citet[][]{spriher}  obtain $C$ in the range $2.5 - 6$ at $z=15$, with higher values corresponding to higher resolutions. Facing those uncertainties, we  adopt  $C = 10$ as a fiducial value at $z=15$. 
 
Let us consider quasars as the sources of the ionising radiation. The typical lifetime  of quasars is not a tightly constrained quantity, and values ranging from $10^6$ to $10^8$ years are usually quoted \citep[e.g.][]{qsolife}. It is nevertheless reasonable to assume that quasars spend an Eddington time scale,
\be\label{t-edd}
 t_\text{E} = 4\times 10^8\ \epsilon\ \text{yrs},
\ee
in their shining phase, and thus live at least that long ($\epsilon\sim 0.1$ is the radiative accretion efficiency). Then, comparing eqs. (\ref{t-rec}) and (\ref{t-edd}), we see  that for $z = 15$,  the recombination time scale is shorter than the quasar lifetime as soon as $C\gtrsim 3.15$. 
Thus, the radius of the ionised bubble can be considered constant. The total time spent by the bubble in its expanding phase is simply given by the ratio of the recombination time scale to the source lifetime. Considering $t_\text{E}$ as the minimum lifetime, this ratio is at most of order
\be
\frac{t_\text{rec}}{t_\text{E}} = 0.315 \, \epsilon^{-1}\, C^{-1}\left(\frac{\Omega_b h^2}{0.024}\right)^{-1}\left(\frac{1+z}{16}\right)^{-3} 
\ee
and is less than $13\%$ of the source lifetime at $z=15$ for $t_\text{S}\sim 10^{8}$ yrs and $C=10$. In the following, we therefore adopt the definition of $r_\text{S}$ given by eq. (\ref{sr}).

Whichever the photon source, the most relevant photons are those that are roughly at the hydrogen ionisation threshold. Indeed, for a stellar source, the UV spectrum is approximately  flat with a cutoff at only a few times the Lyman limit frequency. Similarly, quasar spectra invariably exhibit the so-called Blue Bump at  $\lambda \sim 1100$\AA\ \citep[e.g.][]{telfer, shang}. Let us then count as ionising photons those with energies in the range 13.6 eV - 29.3 eV (ionisation by photons with higher energies is more than ten times less likely to happen). The ionising photons emission rate can be then estimated as
\be
\dot{N}_\text{ph} \simeq \frac{L_{E_0}}{E_0} \simeq \frac{L}{E_0} \frac{\int_{E_0}^{29.3}\phi(r, E) dE}{\int_{E_0}^{E_c} \phi(r, E) dE}
\ee
where $E_0 = 13.6\ \text{eV}$, $L$ is the source luminosity and $E_c$
is the  energy at the spectrum cut-off. For quasars, we have 
\be
\phi(r, E) = \phi_0(r)(E/E_0)^{\alpha} \exp(E/E_c),\label{flux}
\ee
 with $\alpha \sim -1.7$ and $E_c\sim 2$ keV \citep[][]{telfer,sos}, and
\be
\dot{N}_\text{ph} \simeq 0.4\frac{L}{E_0} \simeq 7\times 10^{55} L_{12} \ \text{s}^{-1}
\ee
with $L_{12}=L/ 10^{12}L_\odot$. Notice that all $z>6.1$ quasars observed to date are extremely luminous and their $\dot{N}_\text{ph}$ estimated from observations lie in the range $[0.6 - 2.2] \times 10^{57}\ \text{s}^{-1}$ \citep[see, e.g., Table 1 of][]{yulu}. Finally, we obtain a Str\"omgren radius of
\be\label{stromgren}
r_\text{S} \simeq 683\  C_{10}^{-1/3} L_{12}^{1/3} \left(\frac{\Omega_b h^2}{0.024}\right)^{-2/3} \left(\frac{1+z}{16}\right)^{-2} \ \text{kpc},
\ee
where  $C_{10}=C/10$.

\subsection{Ionised case}
Within the Str\"omgren sphere, the plasma is basically fully ionised, and the dominant mechanism to accelerate electrons is simply the Thomson scattering of photons off free electrons. Therefore,  $\sigma_\gamma \equiv \sigma_{\rm{T}} = 6.65\times 10^{-25}\,\text{cm}^2$, and the source current $I$ in this case is defined by
\be
I \equiv  \frac{E}{m_ec^2}\, q_en_ec.
\ee 
Then, directly from  eq. (\ref{field}), we get
\be
B \sim  \frac{ E \sigma_\text{T} \phi}{R^{\text{rot}}q_e} \,  t_\text{S} 
\ee
where $R^{\text{rot}}$ is the scale of inhomogeneities in the ionising flux $\phi$. 
At the ionisation energy threshold,  we can quickly estimate the relevant photon flux at a distance $r$ from the source by taking
\be
E \phi(r, E) = \frac{L}{4\pi r^2}\label{qf}
\ee
at $E = 13.6$ eV. The amplitude of the field is then
\be
B \sim 4.56\times 10^{-28} \frac{L_{12}}{R^\text{rot}_{2}r_{3}^2} \frac{t_\text{S}}{10^8\ \text{yrs}}\,\,\text{Gauss}\label{ionised}
\ee
where  $R^\text{rot}_{2} = R^\text{rot}/10^2\,\text{kpc}$ and $r_3 = r/10^3\,\text{kpc}$. The field strength we obtain here is
extremely small, and has presumably little relevance for seeding the
protogalactic plasma (see also Sect. \ref{s-term}). Notice that for
small $r$ the amplitude of the field in eq. (\ref{ionised}) could be much higher. However, at short
distances from the source, the plasma is likely to be
turbulent, and outflows perhaps driven by stellar winds, or  jets
powered by central engine activity, may occur. In those situations,
pure plasma instabilities, such as the Weibel instability, are likely
to be much more efficient in producing magnetic fields than the model presented here.

\subsection{Neutral case}\label{nc}
Beyond the Str\"omgren radius, the situation is quite
different. Indeed, the medium is opaque to UV photons at the ionising energy  threshold, and the medium is therefore basically neutral. However, photons with higher energies do propagate further deep into the neutral region and eventually ionise atoms at a distance which depends on the ionisation mean free path.  In this case, the mechanism of electron acceleration is therefore the ionisation process itself.
\begin{figure}[t]
\rotatebox{90}{
\includegraphics[height=85mm]{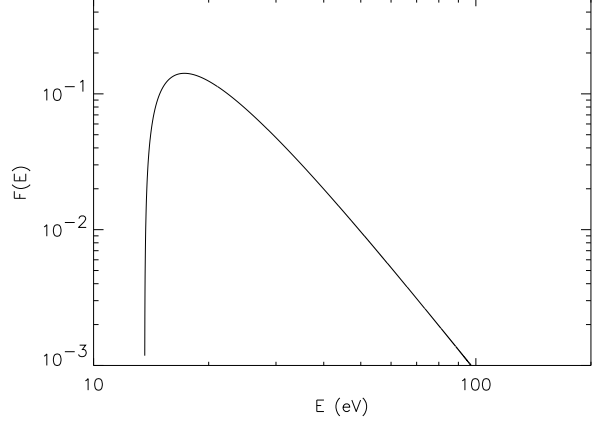}
}
\caption{\label{mte_2} $F(E) = f_\text{mt} (E/E_0)^{-3.7} \exp(E/E_c)$ as a function of the incident photon energy.}
\end{figure}
Thus, $\sigma_\gamma \equiv \sigma_\text{ion}$ is now the ionisation cross-section, which depends on the photon energy as
\be
\sigma_\text{ion} = \sigma_0 \left(\frac{E}{E_0}\right)^{-3}
\ee
with $\sigma_0 \simeq 7.8\times 10^{-18}\,\text{cm}^2$, 
and the source current is now defined by
\be
I \equiv  f_\text{mt}\frac{E}{m_ec^2} q_e n_\text{H} c,
\ee
where $n_\text{H}$ is the density of neutral hydrogen, and $f_\text{mt}$ is the fraction of the incident photon momentum transferred to the photoelectron (notice that we neglected terms of the order of $m_e/m_p$). The amplitude of the seed field is now estimated to be
\be
B \sim f_\text{mt} \frac{n_\text{H}}{n_e} \frac{ E\sigma_\text{ion} \phi}{R^{\text{rot}}q_e}\,   t_\text{S}.
\ee
Using eq. (\ref{flux}) as a definition for the ionising photon flux, 
we normalise the spectrum at a distance $r$ by estimating the power emitted by the source, i.e.
\be
\int_{E_0}^{E_c} \phi(r) dE = \frac{L}{4\pi r^2}
\ee
which gives
\be
\phi_0(r) \simeq 5.4\times 10^{-2} \frac{L}{E_0 r^2}, 
\ee
and in this case, the field amplitude obtained is
\be
B\sim 7.6\times 10^{-16} F(E) \frac{(n_\text{H}/n_e)}{10^4}\frac{L_{12}}{R^\text{rot}_2 r^2_\text{S}}\frac{t_\text{S}}{10^8\ \text{yrs}}\ \text{Gauss},
\ee
where $F(E) =  f_\text{mt} \left(E/E_0\right)^{-3.7}
\exp\left(E/E_c\right)$.
Notice that in the equation above, as compared to eq. (\ref{ionised}), $r_3$ has been replaced by $r_\text{S}/1\,\text{Mpc}$, the Str\"omgren radius normalised to 1 Mpc, since we are interested in the effects of photons penetrating into the quasi-neutral medium outside the H\textsc{ii} bubbles. 
Using the definition of the Str\"omgren radius in eq. (\ref{stromgren}), the amplitude of the field obtained is finally
\bea
B &\sim& 1.63\times 10^{-15} F(E)  C_{10}^{2/3}\frac{(n_\text{H}/n_e)}{10^4} \frac{L_{12}^{1/3}}{R^\text{rot}_2} \frac{t_\text{S}}{10^8\ \text{yrs}} \nn\\
&\times&\left(\frac{\Omega_b h^2}{0.024}\right)^{4/3} \left(\frac{1+z}{16}\right)^4  \,\,\text{Gauss}\label{neutral}.
\eea

\begin{figure}[t]
\includegraphics[width=85mm]{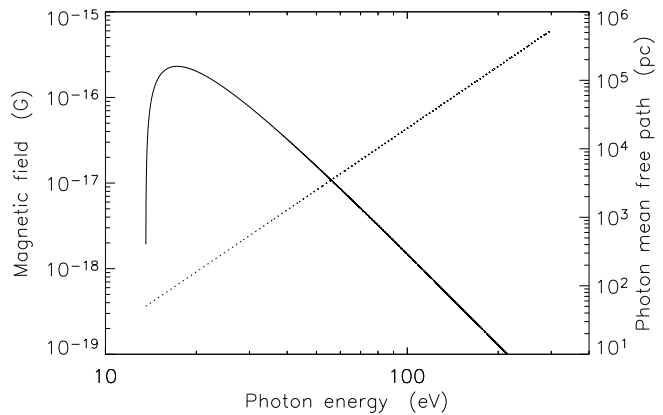}
\caption{\label{amp} Amplitude of the magnetic field (solid line) obtained and photon mean free path (dashed line) as functions of the incident photon energy.}
\end{figure}

\begin{table}[t]
\begin{center}
\begin{tabular}{|c|c|c|c|c|c|}
\hline
\hline
$E$ (eV)& $17.3$ & $25$ & $30$ & $53$ & $136$ \\
\hline
$l_\text{mfp}$ ($10^2$ pc) & $1.03$ & $3.12$ & $5.39$ & $29.7$ & $502$\\
\hline
$B$ ($10^{-17}$ G)  & $23.1$ & $12.7$ & $7.77$ & $1.3$ & $0.05$\\
\hline
\hline
\end{tabular}
\caption{\label{resu} Photon mean free path in the neutral medium as a function of their energy and corresponding values of the generated magnetic field at $z=15$ (assuming a fluctuation scale of $100\ \text{kpc}$ and a clumping factor of $C=10$).
}
\end{center}
\end{table}
Now, the fraction of momentum transferred to a photoelectron depends on the energy of the incident photon \citep[e.g.][]{gm}, 
\be
f_\text{mt} = \frac{8}{5}\frac{E - E_0}{E},
\ee
such that  $f_\text{mt}\simeq0$ at the ionisation threshold, and the amplitude of the field obviously vanishes. However, as shown on Fig. \ref{mte_2} where we plot $F(E)$ as a function of $E$, as soon as we are slightly above the threshold, $B>0$, and the most favorable photons for magnetic field generation have $E \simeq 17.3$ eV. The photon mean free path outside the Str\"omgren sphere is $l_\text{mfp} = (\sigma_\text{ion} n_\text{H})^{-1}$, i.e.
\be\label{lmfp}
l_\text{mfp} =50.2  \left(\frac{E}{E_0}\right)^3 \left(\frac{1+z}{16}\right)^{-3}\left(\frac{\Omega_b h^2}{0.024}\right)^{-1} \ \text{pc}
\ee
and is increasing with  energy. This implies that the strongest fields
are created just within the thin skin of the ionised bubble, whereas
slightly weaker fields are generated on thicker depths. Assuming a
clumping factor of $C=10$, Fig. \ref{amp} shows the amplitude of the magnetic field we obtain (solid line), as well as the photon ionisation mean free path (dashed line), as functions
of the photon energy.  Table \ref{resu} summarises a few typical values of the magnetic field and of the depth on which it is created.  
As can be seen, the amplitude of the field remains above  $10^{-17}$ Gauss on kiloparsec scales and decreases down to $5\times 10^{-19}$ Gauss at 50 kpc. 

These amplitudes depend on the scale $R^\text{rot}$ of the inhomogeneities in the neutral gas surrounding the Str\"omgren sphere. The maximum scale that we can consider is therefore of order $r_\text{S}/2\sim 340$ kpc. On the other hand, from structure formation, we expect fluctuations of matter density on all scales. However, neutral hydrogen fluctuations will be present only in haloes in which the virial temperature is above  $10^4$ K. The comoving number density of such haloes is roughly $20\, h^{3}\text{Mpc}^{-3}$ \citep[][]{mowhite}, which corresponds to a proper mean separation of $\sim 32$ kpc at $z=15$.  Below that temperature threshold, in absence of heavy elements, the primordial gas is unable to cool and condense. Thus, taking $R^{\text{rot}} = 100$ kpc for illustrative purposes in eq. (\ref{neutral}) is fully justified, and the actual value of $B$ may  differ at most by a factor of $\sim 3$.

In addition, the amplitude of the field we obtain depends on the actual ratio
$n_\text{H}/n_e$ at redshift $z=15$. This ratio is essentially given
by the residual ionisation state of the universe from recombination
era.  By taking $n_e/n_\text{H} \sim 10^{-4}$ in eq. (\ref{neutral}), we may have slightly overestimated the recombination efficiency. In fact, recent calculations \citep[][]{sss} including non-equilibrium effects give rather $n_e/n_\text{H} \sim \text{a few}\, 10^{-4}$ \citep[see also][]{hssw}.

Finally, note also that the amplitude of the field is only weakly dependent  on the luminosity of the source, meaning that quasars with $L = 10^{10} - 10^{13}\,L_\odot$ will produce fields of similar strength. This broadens the applicability of the mechanism presented here, in that it is not restricted to the most powerful, and rare,  sources.

\subsection{Source term}\label{s-term}
Examining  eq. (\ref{field}), it is obvious that the magnetic field
will be  non-zero only if the source term has a rotational part. 
This is actually the case if the medium in which photons propagate is inhomogeneous. Indeed, if we decompose the radiation flux into a  mean background $\vec{\phi}_0$ and an inhomogeneous component $\vec{\delta\varphi}$, 
\be
\vec{\phi} = \vec{\phi}_0 +\vec{\delta\varphi} = (1+f_l) \vec{\phi}_0,
\ee
we can relate the fluctuations of the flux to the fluctuations of the optical depth $\tau_l$ according to 
\be
f_l = e^{-\tau_l} -1 \label{opt1}
\ee
and thus also to the matter density inhomogeneities present in the plasma, since
\be
\tau_l = \sigma_{\gamma} \bar{n}_e\int^l_0\delta(\vec{r}) dr_\pa \label{opt2}
\ee
where $\bar{n}_e$ is the mean electron number density, $l$ is the photon propagation path, and $\delta(\vec{r}) = (n_e(\vec{r}) - \bar{n}_e)/ \bar{n}_e$ is the density contrast. Finally, in the optically thin limit, 
\be
f_l\simeq - \sigma_{\gamma} \bar{n}_e \int^l_0\delta(\vec{r}) dr_\pa \label{opthin},
\ee
and thus
\be
\rot \vec{I} \propto \left(\vec{\nabla}f_l \times \vec{\phi}_0 + f_l\rot \vec{\phi}_0\right).
\ee
Obviously, since  $\vec{\phi}_0$  depends only on $|\vec{r}|$, the
second term above vanishes. The first term however, has no reason to
be zero, since at a given distance $l$ outside the Str\"omgren sphere,
the neutral medium is inhomogeneous, and the optical depth has a non
trivial angular dependence. Inside the ionised bubble, on the
contrary, the ionisation process has probably washed out most of the
density fluctuations. These effects add to the tiny amplitude obtained in eq. (\ref{ionised}), meaning that radiation drag most probably does not create significant fields within the Str\"omgren sphere.

\section{Discussion}\label{fin}
In \citet[][]{first}, we presented a first mechanism for the generation of magnetic fields. In that paper,  we explored, in the steady state,  the effects of radiation drag  provided by ionising sources at the end of reionisation, i.e. when the inhomogeneous universe is fully ionised. The amplitudes that could be reached in that case are as high as $10^{-12}$ Gauss on protogalactic scales. Here, we considered the early stages of inhomogeneous reionisation. The main difference in this situation is that the dominant interaction between electrons and photons is photoionisation, of which the cross-section is orders of magnitude higher, in a suitable photon energy range, than that of Thomson scattering (which dominates in the fully ionised case). Therefore, we might have expected to get even higher magnetic field amplitudes than in the ionised case,  albeit  the  momentum fraction transferred from photons to electrons is smaller than one for photon energies below 36.3 eV.
However, the amplitudes we obtain here are in fact not higher than in the fully ionised case. This is mainly due to the combined effects of a finite source lifetime and of the early dynamical importance of nonlinear terms.  

For initially vanishing fields, nonlinear effects in eq. (\ref{ohms}) will not alter the generation scenario outlined above. They will, however,  play a major role for the field amplification once a critical value is reached. Considering the fluid velocity field due to structure formation, it is easy to show that the first nonlinear term to enter into play in the generalised Ohm's law is $\vec{u}\times\vec{B}$.

In the linear regime, the structure formation process  creates a velocity field 
\be
\vec{u} = \frac{2}{3}\frac{f(\Omega)}{H\Omega}\vec{g}\,\ \ \text{with}\ \ \div \vec{g} = 4\pi G \rho\delta
\ee
with $f(\Omega)\simeq \Omega^{0.6}$. This term enters into play in the usual induction equation for the magnetic field as 
\be
\pd_t\vec{B} = \rot \left(\vec{u}\times \vec{B}\right) + 4\pi\frac{\nu_\gamma c}{\omega_p^2}\rot \vec{I}
\ee
and dominates over the source term as soon as $B\sim 4\pi (c/u) (\nu_\gamma/\omega_p^2) I$, i.e. at a time $t_\text{nl}\sim R^{\text{rot}}/u$ (from eq. \ref{field}). On a given scale, the velocity field is best described by its Fourier mode
\be
|\vec{u}_k| = \frac{H f(\Omega)}{k}\delta_k,
\ee 
and for $k\sim 1 / R^{\text{rot}}$, 
\be
t_\text{nl}\sim \left(H f(\Omega) \delta_k\right)^{-1} \simeq 4.1 \times 10^8\ \delta_k^{-1}\ \text{yrs}
\ee
at $z=15$ with $\Omega = 1$, $\Omega_m=0.27$, and $h=0.71$. This shows that nonlinear terms enter into play basically immediately after the period of magnetic field generation. It implies that magnetic field amplification by shear and adiabatic contraction sets-in directly from the seeds calculated in Sect. \ref{nc}, without any decrease in amplitude by diffusion or dilution due to the expansion of the universe.

Finally, as seen from eq. (\ref{lmfp})  photons with higher energies 
 propagate further, so it would be tempting to calculate the magnetic field produced by photons with $E \sim 400$ eV for instance. Notice however that photons with an energy $E\gtrsim 350$ eV travel freely distances longer than 880 kpc, which corresponds to the mean proper distance between those rare sources which are embedded in $3.9\sigma$ dark matter haloes at $z = 15$ \citep[e.g.][]{mowhite}. Those photons therefore do not contribute to the generation of magnetic fields on such distances, as there are no more privileged directions along which electric fields could be generated, but they rather contribute to establish a uniform soft X-ray background which partially ionises the intergalactic medium \citep[][]{aparna,oh, madau}. Nevertheless, photons with energies $\sim 280$ eV, travelling roughly half the mean source separation, could lead to the creation of  magnetic fields with amplitudes of the order of $4\times 10^{-20}$ Gauss, premagnetising therefore the intergalactic medium at a redshift of $z=15$.

\section{Conclusion}\label{fin2}
We reported here on a new model for the generation of cosmological magnetic fields. We extended the mechanism of charge acceleration by radiation drag, which we first explored in a previous article  in the stationary regime, at the end of reionisation \citep[][]{first}. In the present paper, we considered  the transient regime in the case of a neutral, inhomogeneous universe at the onset of reionisation, at a redshift of $z=15$. The driving process is the photoionisation itself which leads to the creation of magnetic fields around the Str\"omgren spheres produced by the ionising sources. In a source lifetime, the magnetic field amplitudes we obtain are in the range $2\times 10^{-16} - 4\times 10^{-20}$ Gauss on $100$ parsecs to $880$ kiloparsecs, i.e. roughly from the Str\"omgren sphere skin to the mean distance between sources at $z=15$. The growth of the amplitude is limited by the source lifetime $t_\text{S}$, but on the other hand, nonlinear terms, driven by velocity fields due to structure formation, become of significant importance at about  $t_\text{S}$. This means that the fields begin their  amplification by shear and adiabatic contraction directly, without decrease in amplitude due to diffusion or by the expansion of the universe.

\begin{acknowledgements}
The work of ML is supported by a Marie Curie Individual Fellowship of the European Community ``Human Potential'' programme, contract number MCFI-2002-00878.
Partial support from the the French space agency CNES, as well as from the French Ministry of Research through the programme {\em ACI - Jeunes Chercheurs}, ``De la physique des hautes \'energies \`a la cosmologie observationnelle'',  is also acknowledged.
\end{acknowledgements}

\bibliographystyle{aa}
\bibliography{Mag3}

\begin{thebibliography}{34}
\expandafter\ifx\csname natexlab\endcsname\relax\def\natexlab#1{#1}\fi

\bibitem[{{Beck} {et~al.}(1996){Beck}, {Brandenburg}, {Moss}, {Shukurov}, \&
  {Sokoloff}}]{becketal}
{Beck}, R., {Brandenburg}, A., {Moss}, D., {Shukurov}, A., \& {Sokoloff}, D.
  1996, \araa, 34, 155

\bibitem[{{Benson} {et~al.}(2001){Benson}, {Nusser}, {Sugiyama}, \&
  {Lacey}}]{bnsl}
{Benson}, A.~J., {Nusser}, A., {Sugiyama}, N., \& {Lacey}, C.~G. 2001, \mnras,
  320, 153

\bibitem[{{Biermann}(1950)}]{battery}
{Biermann}, L. 1950, Zs. Naturforsh., 5a, 65+

\bibitem[{{Carilli} \& {Taylor}(2002)}]{CaTa}
{Carilli}, C.~L. \& {Taylor}, G.~B. 2002, \araa, 40, 319

\bibitem[{{Davies} \& {Widrow}(2000)}]{dw}
{Davies}, G. \& {Widrow}, L.~M. 2000, \apj, 540, 755

\bibitem[{{Giovannini}(2004)}]{massimo}
{Giovannini}, M. 2004, International Journal of Modern Physics D, 13, 391

\bibitem[{{Gnedin} {et~al.}(2000){Gnedin}, {Ferrara}, \& {Zweibel}}]{gfz}
{Gnedin}, N.~Y., {Ferrara}, A., \& {Zweibel}, E.~G. 2000, \apj, 539, 505

\bibitem[{{Gnedin} \& {Ostriker}(1997)}]{go}
{Gnedin}, N.~Y. \& {Ostriker}, J.~P. 1997, \apj, 486, 581

\bibitem[{{Grasso} \& {Rubinstein}(2001)}]{graro}
{Grasso}, D. \& {Rubinstein}, H.~R. 2001, Physics Reports, 348, 163

\bibitem[{{Hu} {et~al.}(1995){Hu}, {Scott}, {Sugiyama}, \& {White}}]{hssw}
{Hu}, W., {Scott}, D., {Sugiyama}, N., \& {White}, M. 1995, \prd, 52, 5498

\bibitem[{{Hummer} \& {Seaton}(1963)}]{humsea}
{Hummer}, D.~G. \& {Seaton}, M.~J. 1963, \mnras, 125, 437

\bibitem[{{Kim} {et~al.}(1996){Kim}, {Olinto}, \& {Rosner}}]{kor}
{Kim}, E., {Olinto}, A.~V., \& {Rosner}, R. 1996, \apj, 468, 28

\bibitem[{{Kronberg}(1994)}]{kron94}
{Kronberg}, P.~P. 1994, Reports of Progress in Physics, 57, 325

\bibitem[{{Kronberg}(2001)}]{kron}
{Kronberg}, P.~P. 2001, in High Energy Gamma-Ray Astronomy, 451+

\bibitem[{{Langer} {et~al.}(2003){Langer}, {Puget}, \& {Aghanim}}]{first}
{Langer}, M., {Puget}, J., \& {Aghanim}, N. 2003, \prd, 67, 043505

\bibitem[{{Madau} {et~al.}(1999){Madau}, {Haardt}, \& {Rees}}]{mhr}
{Madau}, P., {Haardt}, F., \& {Rees}, M.~J. 1999, \apj, 514, 648

\bibitem[{{Madau} {et~al.}(2004){Madau}, {Rees}, {Volonteri}, {Haardt}, \&
  {Oh}}]{madau}
{Madau}, P., {Rees}, M.~J., {Volonteri}, M., {Haardt}, F., \& {Oh}, S.~P. 2004,
  \apj, 604, 484

\bibitem[{{Martini}(2004)}]{qsolife}
{Martini}, P. 2004, in Coevolution of Black Holes and Galaxies, 170--+,
  astro-ph/0304009

\bibitem[{{Massacrier}(1996)}]{gm}
{Massacrier}, G. 1996, \aap, 309, 979

\bibitem[{{Mo} \& {White}(2002)}]{mowhite}
{Mo}, H.~J. \& {White}, S.~D.~M. 2002, \mnras, 336, 112

\bibitem[{{Oh}(2001)}]{oh}
{Oh}, S.~P. 2001, \apj, 553, 499

\bibitem[{{Peebles} {et~al.}(2000){Peebles}, {Seager}, \& {Hu}}]{psh}
{Peebles}, P.~J.~E., {Seager}, S., \& {Hu}, W. 2000, \apjl, 539, L1

\bibitem[{{Pudritz} \& {Silk}(1989)}]{joe}
{Pudritz}, R.~E. \& {Silk}, J. 1989, \apj, 342, 650

\bibitem[{{Sazonov} {et~al.}(2004){Sazonov}, {Ostriker}, \& {Sunyaev}}]{sos}
{Sazonov}, S.~Y., {Ostriker}, J.~P., \& {Sunyaev}, R.~A. 2004, \mnras, 347, 144

\bibitem[{{Seager} {et~al.}(2000){Seager}, {Sasselov}, \& {Scott}}]{sss}
{Seager}, S., {Sasselov}, D.~D., \& {Scott}, D. 2000, \apjs, 128, 407

\bibitem[{{Shang} {et~al.}(2005){Shang}, {Brotherton}, {Green}, {Kriss},
  {Scott}, {Quijano}, {Blaes}, {Hubeny}, {Hutchings}, {Kaiser}, {Koratkar},
  {Oegerle}, \& {Zheng}}]{shang}
{Shang}, Z., {Brotherton}, M.~S., {Green}, R.~F., {et~al.} 2005, \apj, 619, 41

\bibitem[{{Springel} \& {Hernquist}(2003)}]{spriher}
{Springel}, V. \& {Hernquist}, L. 2003, \mnras, 339, 312

\bibitem[{{Subramanian} {et~al.}(1994){Subramanian}, {Narasimha}, \&
  {Chitre}}]{subra}
{Subramanian}, K., {Narasimha}, D., \& {Chitre}, S.~M. 1994, \mnras, 271, L15

\bibitem[{{Telfer} {et~al.}(2002){Telfer}, {Zheng}, {Kriss}, \&
  {Davidsen}}]{telfer}
{Telfer}, R.~C., {Zheng}, W., {Kriss}, G.~A., \& {Davidsen}, A.~F. 2002, \apj,
  565, 773

\bibitem[{{Valageas} \& {Silk}(1999)}]{patjoe}
{Valageas}, P. \& {Silk}, J. 1999, \aap, 347, 1

\bibitem[{{Venkatesan} {et~al.}(2001){Venkatesan}, {Giroux}, \&
  {Shull}}]{aparna}
{Venkatesan}, A., {Giroux}, M.~L., \& {Shull}, J.~M. 2001, \apj, 563, 1

\bibitem[{{Wasserman}(1978)}]{wasser}
{Wasserman}, I. 1978, \apj, 224, 337

\bibitem[{{Widrow}(2002)}]{widrow}
{Widrow}, L.~M. 2002, Reviews of Modern Physics, 74, 775

\bibitem[{{Yu} \& {Lu}(2005)}]{yulu}
{Yu}, Q. \& {Lu}, Y. 2005, \apj, 620, 31

\end{thebibliography}
\end{document}